\documentclass[reprint,amsmath,amssymb,aps,twocolumn,showpacs,
superscriptaddress,nofootinbib]{revtex4-1}

\usepackage{epsfig,amssymb,amsmath,latexsym}
\usepackage{pifont}
\usepackage{hyperref}

\usepackage{color}
\definecolor{nicered}{rgb}{0.7,0.1,0.1}
\definecolor{nicegreen}{rgb}{0.1,0.5,0.1}

\newcommand{\be}{\begin{equation}}
\newcommand{\ee}{\end{equation}}
\newcommand{\bea}{\begin{eqnarray}}
\newcommand{\eea}{\end{eqnarray}}
\newcommand{\beal}{\begin{aligned}}
\newcommand{\eeal}{\end{aligned}}

\newcommand{\nb}{\nonumber}

\newcommand{\de}{\partial}

\renewcommand\l{\lambda}
\renewcommand\L{\Lambda}

\newcommand\m{\mu}
\newcommand\n{\nu}

\newcommand\s{\sigma}

\renewcommand\l{\ensuremath{\lambda}}

\newcommand\ba{\begin{array}}
\newcommand\ea{\end{array}}

\renewcommand\de{\partial}
\renewcommand\sin{\text{sin}}
\renewcommand\cos{\text{cos}}
\renewcommand\log{\text{log}}

\newcommand\SEC[1]{\medskip\noindent{\sl\bfseries #1}}

\begin{document}

\title{Rotating Black Holes in Higher Order Gravity}

\author{Christos Charmousis}
\affiliation{Laboratoire de Physique Th\'eorique, CNRS, Univ.\ Paris-Sud, 
Universit\'e Paris-Saclay, 91405 Orsay, France}
\author{Marco Crisostomi} 
\affiliation{Laboratoire de Physique Th\'eorique, CNRS, Univ.\ Paris-Sud, 
Universit\'e Paris-Saclay, 91405 Orsay, France}
\affiliation{Institut de physique th\'eorique, Univ.\ Paris Saclay, CEA, 
CNRS, 91191 Gif-sur-Yvette, France}
\affiliation{AIM, CEA, CNRS, Univ.\ Paris-Saclay, Univ.\ Paris Diderot,
Sorbonne Paris Cit\'e, F-91191 Gif-sur-Yvette, France}
\author{Ruth Gregory}
\affiliation{Centre for Particle Theory, Durham University, South Road,
Durham, DH1 3LE, UK}
\affiliation{Perimeter Institute, 31 Caroline St., Waterloo, Ontario,
N2L 2Y5, Canada}
\author{Nikolaos Stergioulas}
\affiliation{Department of Physics, Aristotle University of Thessaloniki, 
54124 Thessaloniki, Greece}
\date{\small \today}

\begin{abstract}
We develop a new technique for finding black hole solutions in modified 
gravity that have ``stealth'' hair, i.e., hair whose only gravitational effect
is to tune the cosmological constant. We consider scalar-tensor theories 
in which gravitational waves propagate at the speed of light, and show that 
Einstein metrics can be painted with stealth hair provided there exists a
family of geodesics always normal to spacelike surfaces. We also present
a novel scalar-dressed rotating black hole that has finite scalar field
at both the black hole and cosmological event horizons.
\end{abstract}


\maketitle

Black Holes are a cornerstone in the study of General Relativity (GR), 
be it theoretical or practical.
From the theoretical perspective, we have several exact solutions 
corresponding to black holes in GR, with charges such as mass, angular 
momentum, and even acceleration or Taub NUT parameters. 
Of these, the most important is the Kerr solution
that describes the rotating black hole; most astrophysical 
black holes are believed to be rotating, indeed, the first
detection of gravitational waves was from the merger of two 
spinning Kerr black holes \cite{Abbott:2016blz}. 

Another key objective in gravity is to explain our universe. One of
the main challenges in cosmology is to explain late time acceleration:
Is it due to dark energy, modified gravity, or a very 
small cosmological constant~$\Lambda$, finely tuned by some as 
yet undiscovered mechanism? One fruitful means of explaining 
the small late time acceleration is to modify gravity in such a way as
to induce (or tune) a cosmological constant, but if gravity is modified,
then it is vital to understand how these modifications affect black holes. 
If it is not possible to construct astrophysically realistic black hole 
solutions, then the theory cannot be considered viable. 

In GR, the Kerr solution (see \cite{Carter:1968ks,CARTER1968399}
for inclusion of a cosmological constant) describes the most general 
axisymmetric stationary rotating black hole, moreover, the event horizon 
telescope \cite{Doeleman:2009te} hopes soon to be able to directly 
image the shadow of the black hole at the centre of our galaxy, possibly 
setting constraints on deviations
from Kerr \cite{Johannsen:2016vqy}. It is important therefore to investigate 
whether spinning black hole solutions exist in modified gravity, and if so, 
do they carry non-trivial extra degrees of freedom?  

In this letter, we focus on a particular family of modified scalar-tensor 
theories of gravity, developing a new technique to find, for the first time, 
astrophysically realistic rotating black hole solutions 
with a nontrivial scalar field, or \emph{stealth hair}.
Our method is based on the Hamilton-Jacobi approach
to finding geodesics. In brief, we prove that a GR solution can also
be a solution to modified gravity if the scalar field is the Hamilton-Jacobi
potential for a geodesic congruence in the spacetime in question.
This allows for a wide range of physically interesting stealth solutions,
in particular we present examples of rotating black holes with stealth hair.

We focus for definiteness on shift-symmetric scalar-tensor theories 
of gravity in the family of Degenerate Higher Order Scalar-Tensor (DHOST) 
theories \cite{Langlois:2015cwa,Crisostomi:2016czh,BenAchour:2016fzp}, 
although the technique can be generalised to 
other modifications of gravity. In particular we will focus on the subset of 
theories where gravitational waves propagate at the speed of 
light,  $c_T=1$, in accord with the recent multi-messenger neutron 
star binary merger observation~\cite{GBM:2017lvd}. 

The most general shift-symmetric scalar-tensor theory of gravity in
which gravitational waves propagate with the speed of light is
\be
\beal
{\cal L}\, &= \, K(X) + G(X) R 
+ A_3 \phi^{\mu} \phi_{\mu \nu} \phi^{\nu} \Box \phi \\
&+ A_4  \phi^{\mu} \phi_{\mu \rho} \phi^{\rho \nu} \phi_{\nu} 
+ A_5  (\phi^{\mu} \phi_{\mu \nu} \phi^{\nu})^2 \,, 
\eeal
\label{ESTlag}
\ee
where $K$, $G$ and $A_i$ are all functions of $X = (\partial\phi)^2$, 
and we abbreviate $\de_\m \phi$ as $\phi_\mu$, and  $\nabla_\n \nabla_\m \phi$
as $\phi_{\m\n}$. In order to propagate a single scalar degree of freedom 
and avoid Ostrogradski instabilities, $A_{4,5}$ are constrained by
\be
\beal
A_4&= -A_3 +\frac{1}{8 G}(4 G_X + A_3 X)(12 G_X + A_3 X) \,, \\
A_5 &= \frac{A_3}{2 G}(4 G_X + A_3 X)\,,
\eeal
\ee
where $G_{X} = \de G / \de X$. These constraints reduce the number of free 
functions to three ($K, G$ and $A_3$) and for simplicity we do not consider 
the cubic Horndeski term in our Lagrangian~(\ref{ESTlag}), though this can 
easily be reinstated if required.

A family of exact static spherically symmetric solutions, with the scalar field 
playing the role of dark energy, was initially found in a class of shift-symmetric 
Horndeski theories \cite{Babichev:2013cya}. It has the nice feature of locally 
describing a Schwarzschild geometry while asymptotically approaching a 
self-tuned accelerating cosmology. Since these solutions acquire a metric 
similar to that of GR while having a non trivial scalar field, they have been 
widely called {\it stealth} solutions\footnote{These solutions were extended 
and generalised in different modified gravity theories with similar properties, 
see for example \cite{Rinaldi:2012vy,Kobayashi:2014eva,Babichev:2017guv,
Chagoya:2016aar,Babichev:2017rti}.}. 
They can be mapped via disformal transformations to stealth solutions 
of DHOST theories with unitary speed of gravitational waves 
\cite{Babichev:2017lmw}, and are free of ghost and gradient instabilities 
\cite{Babichev:2018uiw}. For spherically symmetric stealth solutions in 
DHOST theories see 
\cite{Chagoya:2018lmv,BenAchour:2018dap,Motohashi:2019sen}.

Known stealth solutions are spherically symmetric and all feature 
the same characteristic: a constant kinetic term, $X$, for the scalar field
that does not deform an underlying Einstein geometry.
We suspect that it is this feature of a constant magnitude of $\partial \phi$
that allows stealth hair, thus we look for an Einstein manifold,
$R_{\m\n}= \L g_{\m\n}$, admitting such a solution for $\phi$.
First, the equations of motion under these assumptions 
become 
\be
\beal
&\left[ A_3(X_0) \left({\cal E}_3 - \L X_0 \right) 
- 2 \left(K_X +4 \L G_X\right)|_{_{X_0}} \right] \phi_\m \phi_\n\\
& +\left(K + 2\L G \right)|_{_{X_0}}g_{\m\n} = 0
\eeal
\label{EOMg}
\ee
for the metric and
\be
\beal
&A_3(X_0) \left( {\cal E}_4 +2R_{\m\n\rho\s} \phi^{\n\s}\phi^\m \phi^\rho 
-3 \L X_0 \Box \phi \right) \\
& - 2 \left( K_X +4 \L G_X \right)|_{_{X_0}} \Box \phi =0 
\eeal
\label{EOMphi}
\ee
for the scalar, where for compactness we have defined 
\be
\beal
{\cal E}_3 &\equiv \left( \Box \phi \right)^2 - \left( \phi_{\m\n} \right)^2 \,, \\
{\cal E}_4 &\equiv \left( \Box \phi \right)^3 -3 \Box \phi \left( \phi_{\m\n} \right)^2 
+ 2 \left( \phi_{\m\n} \right)^3 \,.
\eeal
\ee
Setting aside the problem of finding a solution for $\phi$ momentarily,
note that in the above, $A_3$, $G$ etc.\ are all constants,
evaluated at some $X=X_0=(\nabla\phi)^2$. Since $\phi_\mu$ itself
is not necessarily a constant vector, to satisfy the above in general 
we must choose subspaces of the general parameter space for
$\{A_3,G,K\}$. Starting with the top line of \eqref{EOMphi}, we
deduce $A_3(X_0)=0$ (unless spacetime has very special
symmetries as we will see later), hence 
\be
(K_X+4\Lambda G_X)|_{X_0}=0 \,, \label{Gconstr}
\ee
where we emphasise that this is \emph{at the specific value} of $X$,
$X_0$. These two constraints now imply that \eqref{EOMg} is
satisfied, provided we set $\Lambda = -K/(2G) |_{X_0}$. In other words,
the cosmological constant appearing in the Einstein manifold is no
longer the bare cosmological constant included in 
the constant part of the $K$ function. This is the \emph{self-tuning}
property of these gravity theories. 

To sum up: Given a general Lagrangian \eqref{ESTlag}, we first look for 
zeros of $A_3$ that determine the value(s) of $X_0$, then ask that the 
derivatives of $G$ and $K$ are related at $X=X_0$ as required above. The 
effective cosmological constant is then fixed by the ratio of~$K$ to~$G$.

Having established the conditions under which Einstein-like metrics can be
solutions to modified gravity, we now make a key observation that allows 
us to construct a stealth solution on the Einstein manifold: 
Given sufficient symmetry in a spacetime, the geodesic equation
\be
\frac{d^2 x^\m}{d \l^2} + \Gamma^\m_{\rho\s} \frac{d x^\rho}{d \l} 
\frac{d x^\s}{d \l} = 0 \label{geo}
\ee
can be solved using a Hamilton-Jacobi potential $S$, such that the gradient
of the potential gives the tangent vector of the geodesic
\be
\frac{\de S}{\de x^\m} = p_\m =  g_{\mu\nu}
\frac{d x^\n}{d \l}  \,.
\label{HJp}
\ee
Typically, this method is used to simplify the solution of a particular
geodesic (such as the orbit of a planet), however, the form of the potential
can be used over a wider range of co-ordinate values that in the case of
a hypersurface orthogonal geodesic congruence becomes effectively
the whole of the spacetime. Thus, given that $\phi_\mu$ has constant 
magnitude, as does the tangent vector of an affinely parametrised geodesic, 
it is natural to make the identification
\be
\phi \leftrightarrow S \,, 
\label{parallel}
\ee
the properties of the geodesic congruence then will ensure that $\phi$
has the requisite properties to be a stealth solution to the extended
gravity equations of motion. Moreover, this provides a nice physical 
interpretation of the constants appearing in the solution. 

We will now illustrate this technique and find a rotating black hole
with stealth hair. Consider the Kerr-(A)dS geometry \cite{CARTER1968399}
\be
\beal
ds^2 &= - \frac{\Delta_r}{\Xi^2 \rho^2} \left[ dt - a\, \sin^2\theta d \varphi \right]^2 
+ \rho^2 \left( \frac{d r^2}{\Delta_r} + \frac{d \theta^2}{\Delta_\theta} \right) \\
&+ \frac{\Delta_\theta \sin^2\theta}{\Xi^2 \rho^2} 
\left[ a\, dt - \left( r^2 + a^2 \right) d \varphi \right]^2 \,, 
\eeal
\label{Kerr}
\ee
where
\be
\beal
\Delta_r &= \left( 1 - \frac{r^2}{\ell^2}\right) \left( r^2 + a^2 \right) -2Mr \,, \;\;
\Xi = 1 + \frac{a^2}{\ell^2} \,, \\
\Delta_\theta &= 1 + \frac{a^2}{\ell^2}\cos^2\theta \,,\qquad
\rho^2 = r^2 + a^2 \cos^2\theta \,, 
\eeal
\ee
$M$ is the black hole mass, $a$ the angular momentum parameter 
and $\ell=\sqrt{3/\Lambda}$ is the de Sitter radius, related to the 
effective cosmological constant (for AdS reverse the sign of $\ell^2$).

Applying the Hamilton-Jacobi technique, we note that the components of 
the metric \eqref{Kerr} are independent of $t$ and $\varphi$, thus $E=-p_t$ 
and $L_z=p_\varphi$ are two constants of the motion, identified with 
the energy and the azimuthal angular momentum respectively.
A third constant of motion is the magnitude of the tangent vector 
$g^{\m\n}p_\m p_\n=X_0=-m^2$, associated with the rest mass of the
test particle\footnote{Note, for illustration we take timelike geodesics,
should a spacelike congruence be required, substitute $m^2 \to -m^2$
in the derivation.}. Most importantly however, a fourth
constant was discovered by Carter \cite{Carter:1968rr}
(here generalised to include $\Lambda$):
\be
\beal
{\cal Q} &= \Delta_\theta \, p_\theta^2 + m^2 a^2 \cos^2\theta \\
&- \Xi^2 \left[ \left( a \, E - L_z\right)^2 - \frac{\sin^2\theta}{\Delta_\theta} 
\left( a \, E - \frac{L_z}{\sin^2\theta}  \right)^2 \right] \,,
\eeal
\ee
who demonstrated that the geodesic equation was separable. We can
therefore write
\be
S = - E\, t + L_z \varphi + S_r (r) + S_\theta (\theta) \,, \label{S}
\ee
where
\be
S_r = \pm \int \frac{\sqrt{R}}{\Delta_r} dr \,, \qquad 
S_\theta = \pm \int \frac{\sqrt{\Theta}}{\Delta_\theta} d\theta \,,
\label{sdefns}
\ee
with
\begin{figure}
\includegraphics[width=0.4\textwidth]{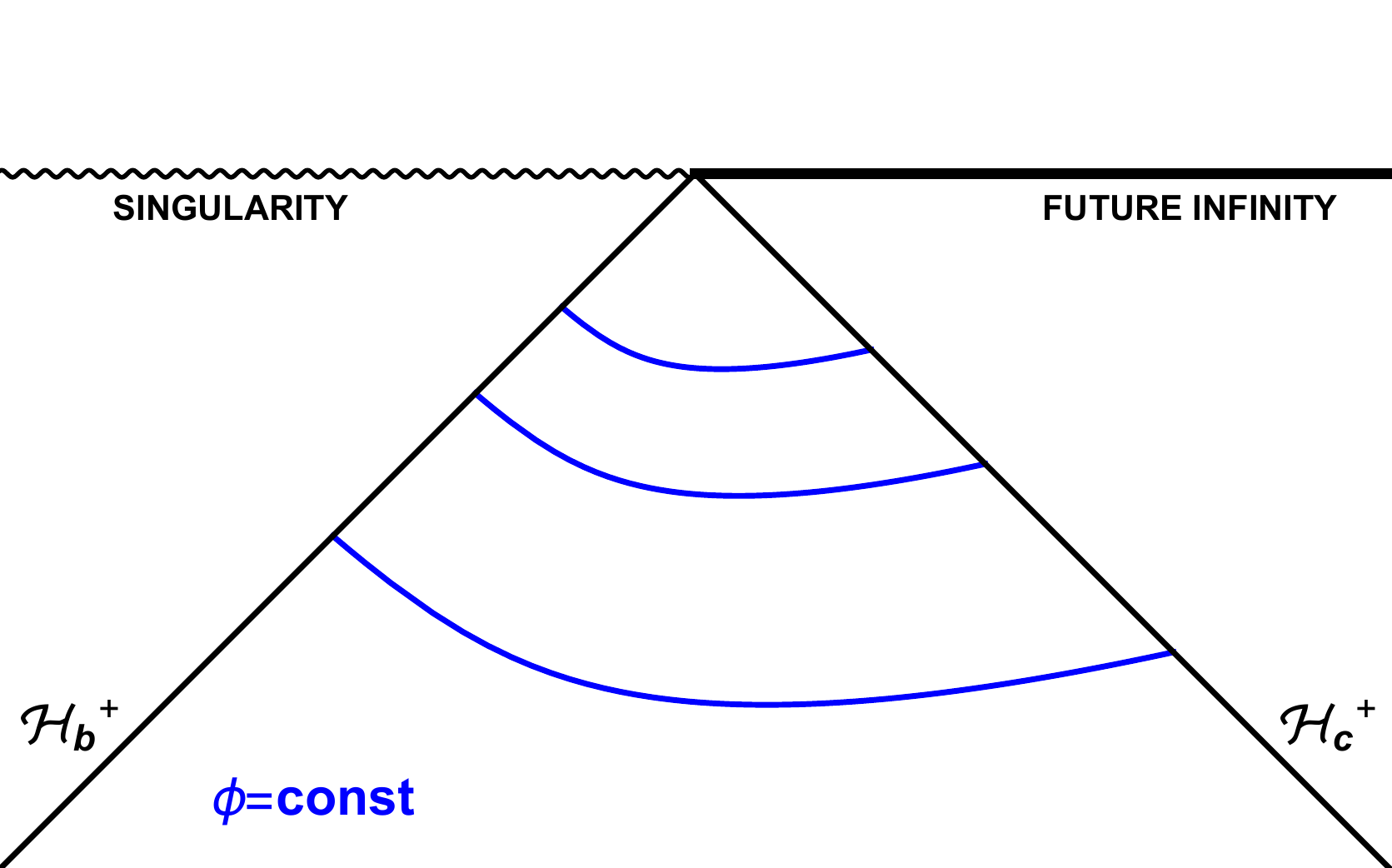} 
\caption{\label{fig:penrose} Contours of constant $\phi$ for 
$a=GM=0.1\ell$, $\eta_c = 0.612$ in local Kruskal coordinates 
for the future event horizons, $\kappa_b U = -e^{-\frac{\kappa_b (t-r^\star)}{2}} ,
 \kappa_cV = - e^{-\frac{\kappa_c  (t+r^\star)}{2}}$, ($\kappa_i$ being
the absolute values of the surface gravities of each horizon).}
\end{figure}
\bea
R &=& \Xi^2 \left[ E \left( r^2 + a^2 \right) -a\,L_z \right]^2 \nb \\ 
&-& \Delta_r \left[ {\cal Q} +  \Xi^2 \left( a \, E - L_z \right)^2 + m^2 r^2  \right] \,, \\
\Theta &=& - \Xi^2 \sin^2\theta \left( a\,E - \frac{L_z}{\sin^2\theta} \right)^2 \nb \\
&+& \Delta_\theta \left[ {\cal Q} +  \Xi^2 \left( a \, E - L_z \right)^2 
- m^2 a^2 \cos^2\theta \right] \,.
\eea

Now let us look for explicit solutions for the scalar field $\phi=S$. This places
further constraints on the potential, as we require $\phi_\mu$ to
be regular throughout the spacetime. Checking regularity on the axes
requires $\partial S/\partial \theta\to0$ as $\theta\to 0,\pi$, i.e.\ 
$\Theta \propto \sin^2\theta$. This in turn requires $L_z=0$ and 
${\cal Q} + \Xi^2 a^2 E^2 = m^2 a^2$, and writing $\Xi E=\eta m$,
we get:
\be
\beal
\Theta &= a^2 m^2 \sin^2\theta \left ( \Delta_\theta - \eta^2 \right) \,, \\
R &= m^2 (r^2+a^2)\left (\eta^2 (r^2+a^2) - \Delta_r\right ) \,.
\eeal
\label{RTheta}
\ee
This has now reduced the parameter space to an overall scaling, $m$,
and a ``relative energy'' $\eta$, constrained to lie in $\eta\in[\eta_c,1]$;
the upper limit coming from $\Theta\geq0$, and the lower limit from 
$R\geq0$ in \eqref{RTheta}.
\begin{figure}
\includegraphics[width=0.35\textwidth]{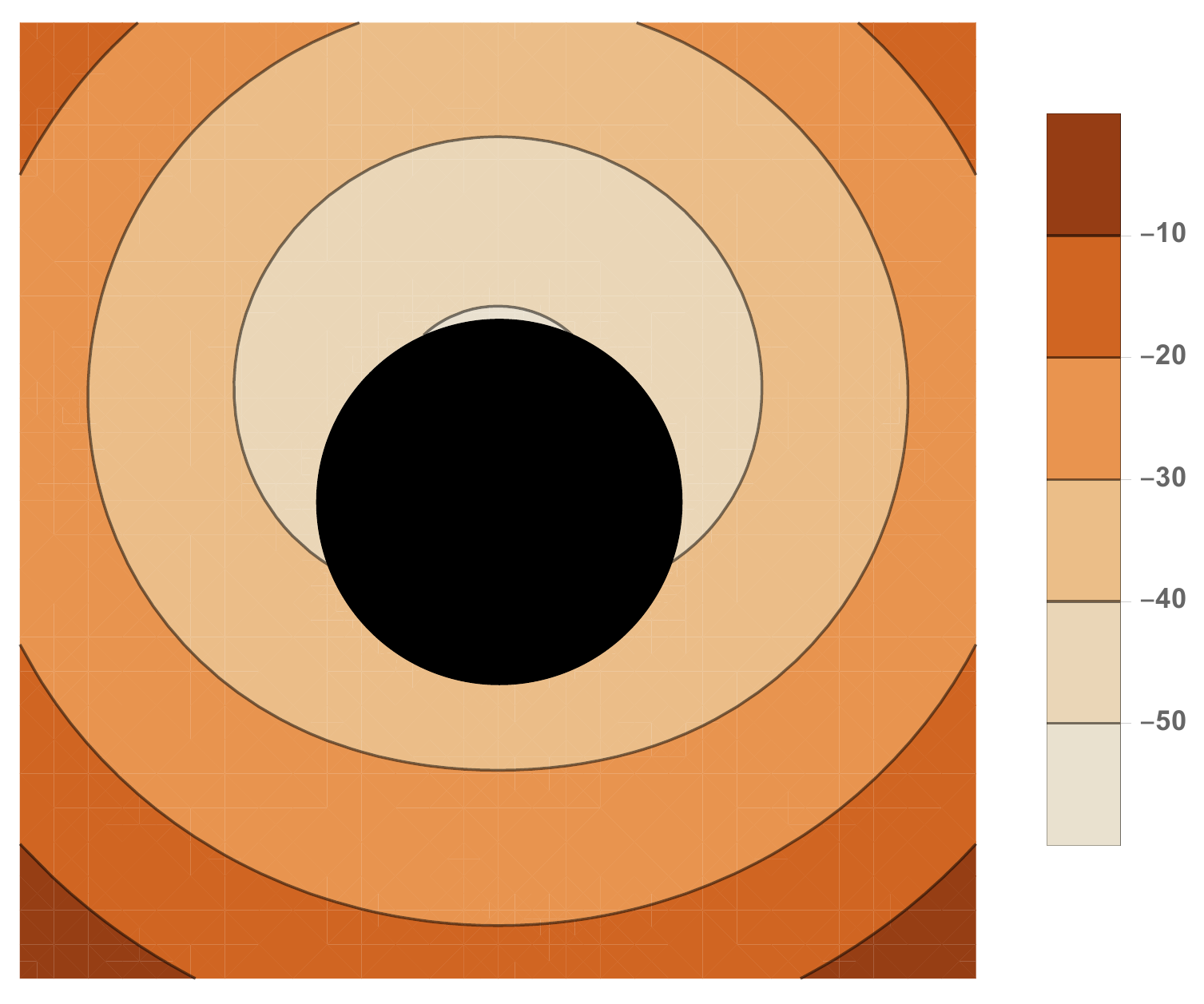} 
\caption{\label{fig:phicont} 
Contours of $\phi$ at constant $v=t+r^\star$ in the $\{r,\theta\}$ plane 
near the black hole horizon with the same parameter values
as figure \ref{fig:penrose}, taking $m=100$.
}
\end{figure}

At first sight, it appears we have four distinct solutions coming from
the choice of signs in \eqref{sdefns},
however, an interesting restriction occurs when $\eta=1$ or $\eta_c$.
In this case $\Theta$ (or $R$) vanishes for some value of $\theta$
(or $r$), and the branch choice changes. This is most easily seen for
$\eta=1$, here $\Theta_1 = \frac{m^2 a^4}{\ell^2} \sin^2\theta \cos^2\theta$,
and the natural root is $\cos\theta$ which changes sign across the 
hemisphere. The same phenomenon occurs for $R$, but this
leads to an important consequence as we now discuss.

Inspection of \eqref{sdefns} shows that 
$S_r \sim \pm m\eta r^\star$ near the event horizons, where
$r^\star = \int dr(r^2+a^2)/\Delta_r$ is the tortoise coordinate,
therefore, if we interpret $\sqrt{R}$ as being the positive root,
our scalar field will be divergent at one or the other horizon
(dependent on the branch choice). 
Note however, that for $\eta_c$, $R$ has a quadratic zero at
some $r_0$: $R\sim R''(r_0) (r-r_0)^2/2$, thus the true root,
$\sqrt{R}\sim(r-r_0)$, changes sign at $r_0$. This means
that for $\phi$ to be differentiable, we must change the sign 
of $\sqrt{R}$ across $r_0$ and set
\be
S_r = (H[r-r_0] - H[r_0-r]) \int_{r_0}^r \frac{|\sqrt{R}|\,dr}{\Delta_r}
\ee
where $H$ is the Heaviside step function.
This now renders $\phi$ finite at both future event horizons,
and infinitely differentiable between the horizons as shown
in figure \ref{fig:penrose}.

It is worth emphasising this last point: All black hole solutions in
the literature for higher order scalar-tensor gravity
are spherically symmetric, and
have scalar fields that diverge either on the black hole or
cosmological event horizon. While this is not a physical
problem when $\phi$ interacts with gravity only through its gradient, 
it is nonetheless a less appealing feature of these
solutions. Here, we have constructed a \emph{rotating}
black hole with \emph{finite} stealth scalar hair. This scalar
will be manifestly continuous across each horizon, and be
straightforward to analyse in perturbation theory.
Finally, the integral for the $\theta-$potential $S_\theta$ gives
\be
\beal
\pm S_\theta = &\eta \,\log \left [
\frac{ \sqrt{1-\eta^2+\frac{a^2}{\ell^2}\cos^2\theta}
+ \frac{a}{\ell} \cos\theta}{\sqrt{(1-\eta^2)\Delta_\theta}} \right]\\
&\;\;\;-\log \left [
\frac{ \sqrt{1-\eta^2+\frac{a^2}{\ell^2}\cos^2\theta}
+ \frac{a}{\ell} \cos\theta}{\sqrt{1-\eta^2}} \right]\;,
\eeal
\ee
leading to an ``off-centre'' behaviour in the scalar as shown in 
figure \ref{fig:phicont}.

\begin{figure}
\includegraphics[width=0.35\textwidth]{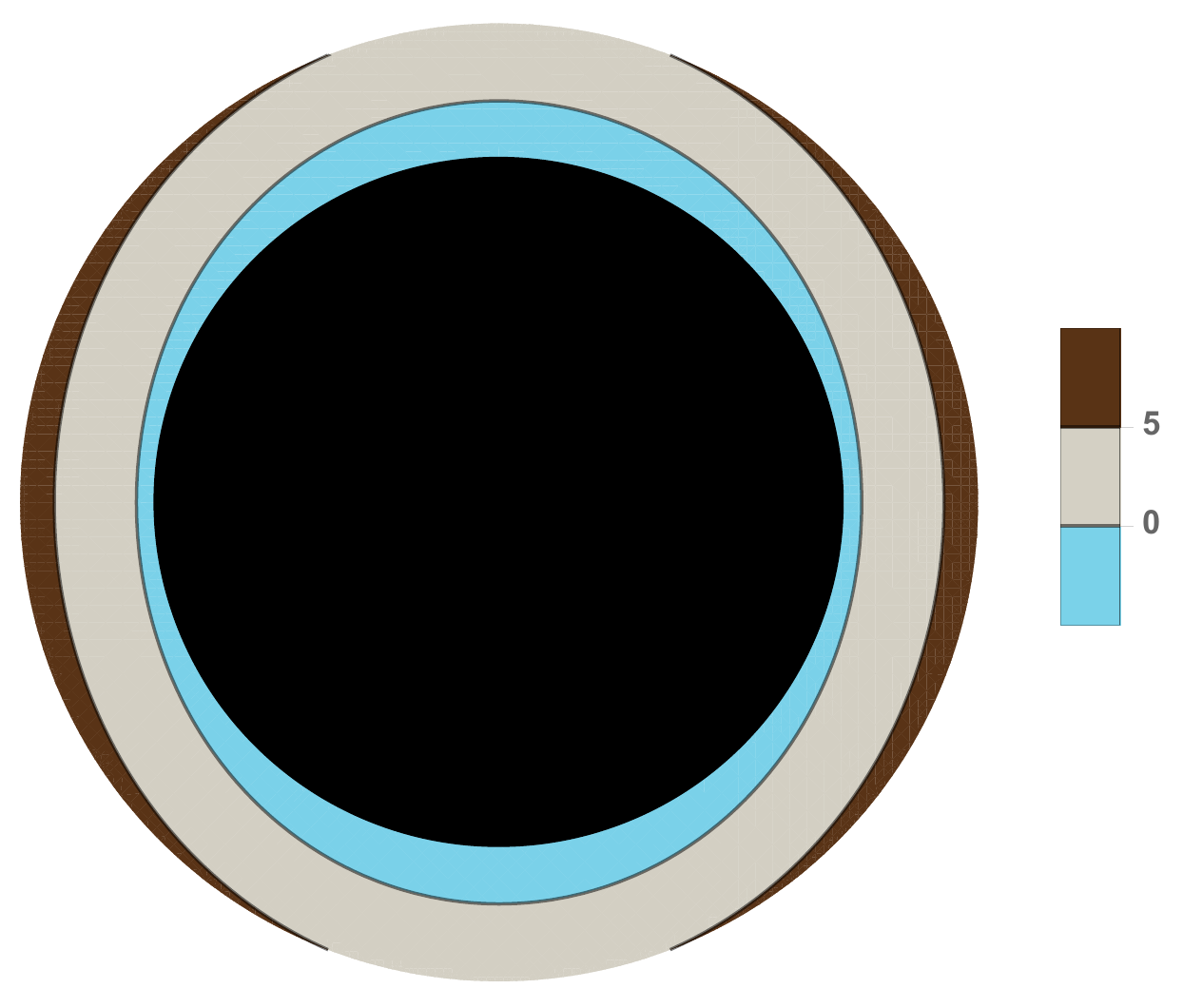} 
\caption{\label{fig:nari} 
Contours of $\phi$ for $\eta=1$ at constant $v=t+r^\star$ in the 
$\{r,\theta\}$ plane. ($a=GM=0.22\ell$, $m=100$).
}
\end{figure}

For $\eta>\eta_c$, the radial function $R$ has no zero, and 
the scalar field diverges on one horizon, in common with the 
known solutions in the literature. The field also demonstrates a 
similar asymmetry in $\theta$, except for $\eta=1$, when 
\be
S_\theta =  \pm \frac{m\ell}{2} \log \Delta_\theta \,,
\ee
rendering the angular variation symmetric about the equator
as shown in figure \ref{fig:nari}.

Our solution for $\phi$ shows a clear dependence on both $\theta$ and $r$,
as well as the time dependence in common with known stealth solutions 
having spherical symmetry \cite{Babichev:2013cya}. The key 
difference is that we can construct a scalar that is finite on
both the black hole and cosmological horizon. To compare to
solutions in the literature, we take spherical symmetry ($a\to0$)
and find general solutions of the form $\phi = -m \eta t \pm S_r$.
Once again, we have a finite-$\phi$ solution for $\eta_c^2 
= 1-3(M/\ell)^{2/3}$, 
\be
\phi = - m \left [ \eta_c t + \int \frac{r(r-r_0) \sqrt{r(r+2r_0)}}{\ell\Delta_r} dr\right]\,,
\label{regso3phi}
\ee
where $r_0 = M^{1/3} \ell^{2/3}$. It is interesting to compare this to
the time-dependent solution of a black hole in slow-roll inflation
\cite{Chadburn:2013mta,Gregory:2018ghc}. 
There, the scalar profile $\phi_{SR}\propto T$ (for a suitable time
coordinate) is also finite at both horizons, but \eqref{regso3phi}
has constant gradient, whereas $\phi_{SR}$ 
solves a wave equation, resulting in a slightly different
radial profile between the horizons.

Also, note that in the case of spherical symmetry, we can relax our 
constraint $A_3(X_0)=0$. In this case ${\cal E}_3, \,{\cal E}_4$ and the 
Riemann tensor term appearing in (\ref{EOMg},\ref{EOMphi}) have a 
simple form, and combine to require
\be
(K_X+4\Lambda G_X+\frac{3}{2}\Lambda X A_3)|_{X_0}=0 \,,
\ee
with the same self-tuning condition for the cosmological 
constant\footnote{See \cite{Babichev:2016kdt} for a precise analysis 
of self tuning conditions in spherical symmetry.}, however, note
that $A_3(X_0)\neq0$ \emph{requires} $\eta\equiv1$, thus we
no longer have the $\eta-$degree of freedom.
What now happens is that spacetime
becomes foliated by surfaces of constant $\phi$ that are
flat (for $\L=0$).

Another interesting possibility for spherical symmetry is a static 
solution, found by setting $E=0$, hence $X_0>0$ and $\phi$
corresponds to a congruence of spacelike geodesics:
\be
\phi_s = S_r(r) = \sqrt{X_0} \int \frac{r}{\sqrt{\Delta_r}}\,dr \,,
\ee
agreeing with a solution reported in \cite{Motohashi:2019sen}.

It is also worth noting a side result of our analysis: A search for 
solutions with $X=X_0$ and spherical symmetry in $c_T=1$ 
theories (\ref{ESTlag}) allows \emph{only} for Einstein geometries. 
This is unlike Horndeski theories (with $c_T\neq 1$) where solutions 
of black holes and solitons were found that have $X=X_0$ but are 
\emph{not} Einstein spaces (for a concise up to date review see 
\cite{Lehebel:2018zga}). This is an interesting feature, hinting that 
solutions of $c_T=1$ theories, where $X$ is not constant belong to 
branches that eventually flow towards $X$ constant solutions with 
an Einstein space metric.

To sum up: we have presented the first exact solutions for a rotating black
hole with scalar hair in shift-symmetric 
scalar-tensor theories with unitary speed for gravitational waves.
Our method was based on an interesting correspondence between 
families of black hole geodesics and stealth solutions in modified gravity. 
The geodesic correspondence can be used to find other stealth solutions, 
for instance for more general type D spacetimes,
or even non-stealth solutions that do not exist in GR --
one only needs to compute the Hamilton-Jacobi potential for a 
geodesic in the relevant spacetime. 
It is also plausible that this technique can be extended to other modifications of 
gravity, such as vector-tensor theories where we know that certain stealth 
solutions exist. Although we did not consider charged black holes, clearly one
can also use this method to find stealth Kerr-Newman solutions.
Perhaps most importantly, we have
presented a scalar solution that is finite at both the cosmological and 
event horizons, thus manifestly extendible beyond the cosmological horizon.
The angular asymmetry of this solution could provide a distinctive
signature of this hair, although this would require a full perturbation 
analysis beyond the scope of this investigation.

\SEC{Acknowledgments.} \;\;
It is a pleasure to thank Eugeny Babichev, Gilles Esposito-Far\`ese, 
Antoine Leh\'ebel, Karim Noui, Eric Gourgoulhon, 
Karim Van Aelst for many interesting discussions. CC and NS would like 
to acknowledge networking support by the GWverse COST Action CA16104, 
\emph{Black holes, gravitational waves and fundamental physics}. 
CC thanks the Laboratory of Astronomy of AUTh in Thessaloniki for 
hospitality during the course of this work.
MC is supported by the Labex P2IO and the Enhanced Eurotalents Fellowship.
RG is supported in part by the STFC [consolidated grant ST/P000371/1], 
and in part by the Perimeter Institute. Research at Perimeter Institute is 
supported by the Government of Canada through the Department of Innovation, 
Science and Economic Development and by the Province of Ontario 
through the Ministry of Research and Innovation.

\end{document}